\documentclass[11pt]{article}
\usepackage{graphicx}
\usepackage{amssymb}
\usepackage{amsmath}
\usepackage{color}
\DeclareMathOperator\arctanh{arctanh}
\usepackage{epstopdf}
 \usepackage{textcomp}  
\usepackage{amsmath,amssymb} 
\usepackage{epsfig,latexsym}
    \usepackage{graphicx}
\newcommand{\sect}[1]{\setcounter{equation}{0}\section{#1}}

\def\al {\alpha}

\def\ba{\begin{eqnarray}}

\def\be{\begin{equation}}
\def\bi{\bibitem}

\def\BB{{\cal B}}

\def\di{\partial}

\def\ea{\end{eqnarray}} 
\def\ee{\end{equation}}
\def\eee{equation~}
\def\eeee{equations~}

\def\fr{\frac}

 \def\gmn{g_{\mu\nu}}

\def\ha{\frac{1}{2}~}

\def\HH{{\cal H}}

\def\in{\infty}

\def\ka{\kappa}

\def\lb{\label} 
\def\ls{\lesssim}

\def\lll{\left(}

\def\La {\Lambda}
\def\LL{{\cal L}}
\def\LLL{\left[}

\def\mn{\mu\nu}

\def\nn{\nonumber}  

\def\nnn{\noindent}

\def\np{\newpage}

\def\Om {\Omega}

\def\OL{\Om_\La}
\def\ra{\rightarrow}

\def\rh{\rho}

\def\rrr{\right)}

\def\RRR {\right]}

\def\sq{\sqrt}
\def\sqg{\sqrt{-g}}
 
\def\sim{\simeq}
\def\sss{spacetime~}
\def\ssss{spacetimes~}

\def\ta{\tau}

\def\ti{\times}

\def\ts{\textstyle}

\def\vs{\vskip 0.5 cm}

\def\1k{\fr{1}{\ka}}

\def\2k{\fr{1}{2\ka}}
\def\2{{\ts\fr{1}{2}}}

\def\3{{\ts{\fr{1}{3}}}}

\begin{document}

%-----------------------------------------TITLE -----------------------------------
\title{{\bf  A note on dark matter and dark energy  
}}
\author{ Joseph Katz\thanks{email:Joseph.Katz@mail.huji.ac.il}
 \\
 \\
 { \it The Racah Institute of Physics}
 \\
{\it  Edmond Safra Campus, Givat Ram, 91904 Jerusalem, Israel}}
\maketitle
  %________________ABSTRACT____________
\begin{abstract} 
\setlength{\baselineskip}{20pt plus2pt}
 \vs
 Since the geometry of our universe seems   to  depend very little on baryonic matter, we consider a variational principle involving only dark matter and dark energy which in addition   make them depend on each other. There are no adjustable parameters  or scalar fields with appropriate equations of state. No quintessence. For a pressure-less, three-flat FRW model, the cosmological ``constant" is now a function of time, positive by definition and always small. It's time derivative or rather its associated     parameter $w$ is always negative and close to minus one. The most interesting point is that the age of the universe and $w$ itself are correlated. Moreover, this   rather unsophisticated model provides a very limited range for both these  quantities and results are in surprising agreement with observed values. The problem of vacuum energy remains what it was;   the problem of coincidence is significantly less annoying.
 \vs\vs\vs\vs
 \nnn PACS numbers 98.80.Jk   ~~ 98.80.Es  ~~ 95.36.+x 
\end{abstract} 
\np

%-------------Section 1 Introduction---------------------
\sect{ Equations and solutions} 
%----------------------- (i) Basics---------------------
\nnn{\it (i) Basics}
\vs
\nnn Planck 2013 results XVI  \cite{Pl}, confirming  previous observations \cite{Ko}, \cite{La}, \cite{Ri}, present us with an image of the universe whose energy is almost entirely   dominated by two unrelated and still mysterious components, dark matter which causes cosmic attraction and dark energy responsible for cosmic repulsion. Baryonic matter is in for less than 5\%.

In ``Searching for insight"  \cite{LB} Lynden Bell wrote {\it  I still have hopes that thoughts based on Mach's Principle may lead us to a definite prediction of the cosmical repulsion.}  This  note is about  such  a Machian taught. The principle is simple,  consequences straightforward with  predictions   well within the limits of observations. Short comings and other comments are  given at the end.

In a 1991 paper, Tseytlin \cite{Ts} presented the following ansatz for a classical low energy effective Action, ``...not a fundamental  Action which should be quantised..." says Tseytlin referring to ``dual-symmetric string theory and consistency  with standard (inflationary) cosmology" to justify this action\footnote{ The same Action was considered by Davidson and Rubin \cite{DR} to show, in particular, that the cosmological constant might be zero.}
\be
S = \fr{\int_\BB \lll  - \1k R+\LL _m\rrr \sqg d^4x}{\int_\BB \sqg d^4x}~~~~,~~~~\ka=\fr{8\pi G}{c^4}.~
\lb{11}
\ee
Notations are standard,  $\LL_m$ is {\it here} the Lagrangian density of dark matter. Baryonic matter is neglected. The boundary of \sss $\BB$ is a closed hypersurface which in Tseytlin englobes the ``the volume of spacetime".  The variational principle applied to this Action provides Einstein's \eeee with a boundary dependent cosmological term that is  self consistently related to the dark matter Lagrangian density:
\be
R^{\mn}-\2 g^{\mn}R=\ka T^{\mn} +g^{\mn}\La~~~{\rm where}~~~
 \La = \2 \ka\fr{\int_\BB \lll  \fr{\di\LL_m}{\di \gmn}\gmn - \LL_m  \rrr \sqg d^4x}{\int_\BB \sqg d^4x}.~
  \lb{12} 
\ee
To demand that $\La$ be a constant amounts to ask for all solutions of   Einstein-de Sitter \eeee  to satisfy an additional global constraint within the boundaries of the domain. This may be a lot to ask. Moreover  for isolated systems  boundaries are usually taken far away from the sources of gravity  but  in   homogeneous cosmological spacetimes the source is everywhere. We shall therefore let $\La$   depend on the boundaries. It is not easy   to see what else  can be   done.

In classical mechanics the Lagrangian is integrated from some starting point to the current time $t$ and the result is varied to get the equations of motion. We shall do the same here choosing the current time to be the cosmic time. The future is not involved. As a result $\La$ will in general depend on time and the  local conservation law  of dark matter will not hold but rather a combination of dark matter and dark energy is conserved.
%-------------------------(ii) The simplest cosmological model---------------------------------- 
\vs
\nnn{\it (ii) The simplest cosmological model}
\vs
\nnn This complicates, of course,
 Einstein's general relativity considerably. However,  application to FRW \ssss is of  relative simplicity. As a matter of illustration we consider a pressure-less 3-flat spacetime with positive dark energy density $\rh$. The dark matter Lagrangian\footnote{See for instance (4.11) in Schutz and Sorkin \cite{SS}.}   $\LL_m= -2\rh$. Take boundaries  at   time $t_1$ and at   $t> t_1$;  \eeee  (\ref{12}) reduce to
\be
3 \fr{\dot a^2}{a^2}=\ka\rh + \La~~,~~\LLL  a^3( \ka\rh + \La) \RRR^{{\bf .}} - \La(a^3)^{{\bf .}}=0~~,~~
 \La(t,t_1)=\ha\fr{\int^{t}_{t_1}\ka\rh  a^3dt}{\int^{t}_{t_1}a^3dt}.~
\lb{13}
\ee
or, in close to standard notations\footnote{In \cite{Pl}, $\Om_m$ and $\OL$ are present day values. Here these quantities vary with time. We shall add a $0$-indice when referring to the present time $t=t_0=0$. Thus at $t_0$, here  $\OL={\OL}_0$, $\Om_m=\Om_{m0}$ and  the age of the universe today $t_U=t_{U0}$.}, 
\be
 \HH^2:=\lll\fr{H}{H_0}\rrr^2
 =\Om_m+\OL~,~\OL = \HH^2+ {\ts\fr{2}{3}}\HH' ,
 \lb{14}
\ee
and 
\be
\OL\!\!\int_{\ta_1}^\ta \al^3d\ta+\int_{\ta_1}^\ta\3\HH' \al^3d\ta=0.
\lb{15}
\ee
$H$ is the expansion rate at any moment, the prime indicates a derivative with respect to $\ta=H_0t$,   ($\ta_0=0$), and
\be
\al:=\fr{a}{a_0}=e^{\int_{0}^\ta \HH d\ta}. 
\lb{16}
\ee
The first of \eeee (\ref{13}) is   standard FRW$\La$ cosmology. The second equation has been considered and discussed, according to \cite{PR}, by Bronstein \cite{Br} in 1933. Equation (\ref{15})  is new  and deserves some attention because it looks rather   like an integral solution of a differential equation with a  starting value $\ta_1$. Notice that if $\rh>0$, $\OL>0$ by definition.
\vs
%-----------------------------------(iii) Properties of the differential equation--------------------------
\nnn{\it (iii) A differential form of the integral equation}
\vs
\nnn Since $\ta_1$ is arbitrary any other arbitrary time $\ta_2$ leads to an equation like this
\be
\OL\!\!\int_{\ta_2}^\ta \al^3d\ta+\int_{\ta_2}^\ta\3\HH' \al^3d\ta=c_1\OL+c_2.~
\lb{17}
\ee
$c_1$ and $c_2$ are  functions of $\ta_2$. One can eliminate those constants with two successive derivations. A third derivative eliminates $\ta_2$ as well as $\al$, thanks to (\ref{16}) $\al'=\al\HH$,  leading to a third order differential \eee for the expansion rate $\HH(\ta)$,
\be
 (\HH^2+\HH')\HH''' -9\HH\HH'\lll   \HH^3+2\HH\HH'+\HH''         \rrr -{\ts\fr{5}{3}}{\HH''}^2+3{\HH'}^3=0.
 \lb{18}
\ee
Solutions of \eee (\ref{18}) are solutions of \eee (\ref{17}) but not necessarily of (\ref{15}). 
In particular,
the point (or points)  at some $\ta=\tau_3$  where $ (\HH^2+\HH')=0$, is obviously     a singularity of (\ref{18}).  At that point either $\HH''=\fr{12}{5}\HH^3$ or $\HH''=3\HH^3$. $\ta=\ta_1$ is also a point where $ (\HH^2+\HH')=0$ but  $\HH''=\fr{12}{5}\HH^3$ only. It is interesting to note that $\HH'''$ at $\ta_3$ with $\HH''=3\HH^3$ is undefined and higher order derivatives at that point depend on the choice of $\HH'''$. This is not the case at $\ta_1$ where $\HH'''= - \fr{324}{35}\HH^4$ and all higher order derivatives are uniquely defined. 

Suppose we have a solution $\HH(\ta)$ of (\ref{17}) which satisfies appropriate initial conditions. A first derivative of that \eee  gives a sort of first integral that defines $c_1(\ta_2)$
\be
c_1(\ta_2)=\int_{\ta_2}^{\ta}\al^3 d\ta + \fr{\HH^2+\HH'}{\OL'}\al^3.~~
\lb{19}
\ee
Inserting this $c_1(\ta_2)$ into (\ref{17}) gives a second first integral
\be
c_2(\ta_2)=\OL(\ta)\LLL     \int_{\ta_2}^{\ta}\al^3 d\ta - c_1(\ta_2)  \RRR +\int_{\ta_2}^\ta\3\HH' \al^3d\ta.
\lb{110}
\ee 
Equations (\ref{19}) and (\ref{110}) provide a test of the quality of the numerical integration: the right hand sides must be $\ta-$independent. Moreover the \eeee confirm that the solution of (\ref{18}) is also solution of (\ref{17}) and they also give a set of functions of $\ta_2$ which leads in principle to the value  of $\ta_1$ since $c_2(\ta_1)=c_2(\ta_1)=0$.

This being said, two analytic properties of  the integral \eee  (\ref{17}), as well as of the differential \eee  (\ref{18}), are readily discovered and of  interest. If at  some $\ta\ra\ta_i$, $\HH\ra A/(\ta-\ta_i)$  and if for   $\ta\ra\in$, $\HH\ra B/\ta $ the asymptotic forms are necessarily these:
\be
\lim_{\ta\ra\ta_i} \HH=\fr{2/3}{\ta-\ta_i}~~~{\rm and }~~~\lim_{\ta\ra\in} \HH=\fr{(2+\sq{3})/3}{\ta}.
\lb{111}
\ee
Thus near the singularity, $a\varpropto (\ta-\ta_i)^{2/3}$, like in  the simplest cosmological model, with or without a cosmological constant, but later on $a\varpropto \ta^{(2+\sq{3})/3}\sim  \ta^{1.244}$ which is quite different. 
\vs
%----------------------------------(v) a first order differential equation
\nnn{\it (iv) A first order differential equation}
\vs
\nnn (\ref{18}) is reducible to a first order differential equation in terms of
\be
z:=\fr{\OL}{\HH^2}~~~{\rm and}~~~u(z):=\fr{dz}{d\log\HH}; ~~~{\rm setting}~~~u':=\fr{du}{dz}~~~,~~~K_\pm:=\pm2\sq{3}-3,~
\lb{112}
\ee
(\ref{18}) becomes
\be
(1-z)(1-3z)uu'+2(2-z)u^2+(1-z)(5-z)u - 2z(K_+ - z)(K_- -z)=0.
\lb{113}
\ee
Also (for the first limit see below)
\be
\lim_{\ta\ra\ta_i}z=0~~~,~~~\lim_{\ta\ra\ta_3}z=\3 ~~~~{\rm and}~~~~\lim_{\ta\ra\in}z=K_\pm\sim0.464.
\lb{114}
\ee
Corresponding to $\ta_3$ here  $z=\3$. Again at $z=\3$, $u$ equals either  $-\fr{4}{15}$  where\footnote{This corresponds to $\HH''=\fr{12}{5}$.}   $u'=\fr{26}{35}$    or $u= - \fr{2}{3}$  where $u'$ is not defined. 

 It is interesting to note that (\ref{113}) has  a  singularity at $(0,0)$ through which {\it all} regular solutions $u(z)$ must pass with a  slope $u'(0)$ that is either  equal to $-2$ or $-3$. The line with a slope $u'(0)= - 3$ does not go through either of the values of $u$ at $z=\fr{1}{3}$ but that with $u'(0)=-2$  goes through $(\fr{1}{3}, - \fr{2}{3})$  and one readily finds the unique regular solution: $u=-2z$. This solution   is {\it not} a solution of the integral equation (\ref{17}) but it is of interest as we shall see because if $u=-2z$, $\OL$ is  a positive constant  say ${\OL}_c$ and
\be
\HH(\ta)=\fr{\sq{{\OL}_c}}{\tanh\LLL \fr{3}{2}\sq{{\OL}_c}(\ta -\ta_i)    \RRR}.~
\lb{115}
\ee
If this holds up to the singular point $\ta=\ta_3$ where   $ (\HH^2+\HH')=0$  we  get
\be
{\OL}_c=\fr{4\lll \arctanh \fr{1}{\sq{3} }   \rrr^2}{9\lll    \ta_3 - \ta_i    \rrr^2}.~
\lb{116}
\ee
It so happens that   
$\OL(\ta)$ for $0\le z\le\3$ or $\ta_i\le\ta\le\ta_3$, is quite close to ${\OL}_c$ as we shall see. 
\vs
%-----------------------------------(iv) w------------- ------------
\nnn{\it (v) w}
\vs
\nnn The parameter $w$   is a measure of the time derivative of the expansion rate. It is defined  in Peebles and Ratra \cite{PR}; in our notations,
 \be
 \OL'= - 3\HH\OL(1+w)~~~{\rm near}~~~\ta_0=0.~
  \lb{117}
 \ee
From what we said about ${\OL}_c$ follows that for $\ta\le\ta_3$,  $w\sim-1$  since $\OL\sim{\OL}_c$. Later on, we may use (\ref{117}) to calculate   $w_0$.
\vs
%-----------------------------------(vi) Initial conditions--------------------------
\nnn{\it (vi) Initial conditions}
\vs
Now for experimental values of some parameters and initial conditions for the equations. At the time of this calculations, we took the cosmological parameters from a Nasa table on the Web\footnote{http://lambda.gsfc.nasa.gov/product/map/dr4/params/lcdm \_sz\_lens\_wmap7\_bao\_h0\_v4.ps.}. They differ slightly from    values given in Planck 2013 \cite{Pl}:
 \be
 t_{U0}\sim13.75 ~ {\rm Gyears} ~,~H_0\sim70.4~{\rm ~kms^{-1}Mpc^{-1}}~,~\Om_{m0}\sim0.272 ~,~{\OL}_0\sim0.728.~
  \lb{118}
 \ee
 From Planck 2013, formula (94a) \cite{Pl}, 
 \be
   - 1.73\ls w_0\ls -0.32.~
\lb{119}
 \ee 
The initial conditions for \eee (\ref{18}), suggested by recent observations, are thus
\be
\HH_0=1~~~,~~~\HH'_0=-0.408~~~,~~~  -1\ls\HH_0''\ls 3.62.~
\lb{120}
\ee
For \eee (\ref{113}), 
\be
z_0=0.728~~~,~~~ -5.36\ls u_0\ls 2.18.~
\lb{121}
\ee
Domains of interest\footnote{For a cosmological ``constant" ${\OL'}_0=0$, $h''_0=1.224$ and $u_0=-1.456$.} are $\ta_i\le\ta\le\in$ and $0\le z\le1$ and especially $\ta_3\le\ta\le0$. The calculated age of the universe will be
\be
t_U=-\ta_i/H_0\sim -\ta_i\ti 13.9~~ {\rm Gyears}~~
\lb{122}
\ee
%-------------Section 2 Results and Comments ---------------------
\sect{Results and comments} 
% ------------------------(i) Numerical Results-----------------------
\nnn{\it (i) Numerical results}
\vs
The time  scale is the inverse of the expansion rate today $1/H_0$. Here is a reminder of the different times encountered, this may be helpful. $\ta_1$ is the arbitrary time introduced  in (\ref{13}) to which we shall come back below. The big bang singularity is at $\ta_i$. The time $\ta_2\ge\ta_i$ is arbitrary and plays no other role than to  verify that solutions of  the differential \eee (\ref{18}) are also solutions of the integral equation (\ref{17}).  $\ta_3$ is a time at which the differential equation (\ref{18}) is singular. For $\ta_i<\ta<\ta_3$  the expansion rate is practically constant. And today is $\ta_0=0$. This being said,
Mathematica uses iteration methods which  deal  quite well with the singularity at $\ta_3$ of \eee (\ref{18}). Unfortunately it does less well with \eee (\ref{113}) at $z=\3$. 
The expansion rate $\HH(\ta)$ does not differ much from  standard FRW$\La$ cosmology as can be seen in Figure 1. 
%--------Figure 1------------------
  \begin{figure}[htbp]
\begin{center}
\includegraphics[width=8cm]{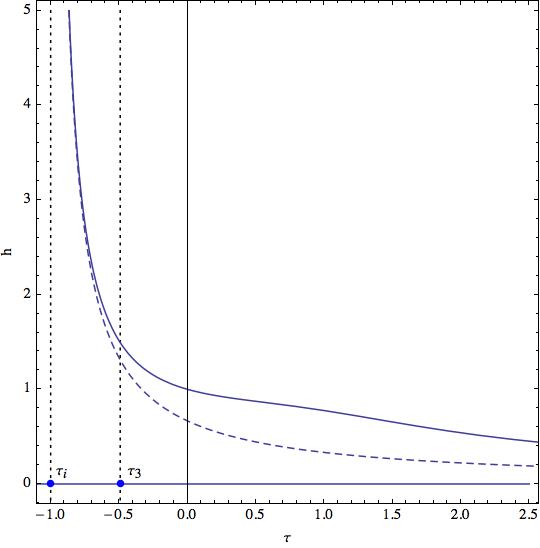}
\caption{A typical example of the   expansion rate or $\HH$ as a function of time $\ta$ for $t_U\sim 13.9$ Gyears   (this is for $\ta_i\sim - 1)$      and $w_0\sim - 0.962$. The dashed curve represents the usual $\fr{2}{3}/(\ta-\ta_i)$   law of standard FRW$\La$  cosmology. The asymptotic continuations follow the laws given in (\ref{111}).    The vertical doted lines are at $\ta_i\sim -1$ and $\ta_3\sim -0.5.$    } 
\label{Figure 1}
\end{center}
\end{figure}
%----------------------
An example of $\OL$ as a function of $\ta$ is shown in figure 2. Notice that  $\OL\sim{\OL}_c>{\OL}_0$, for $\ta\le\ta_3$.  For $\ta>\ta_3$,  $\OL$ decreases   going down to zero at future infinity. 
%------------Figure 2-------------------
 \begin{figure}[htbp]
\begin{center}
\includegraphics[width=8cm]{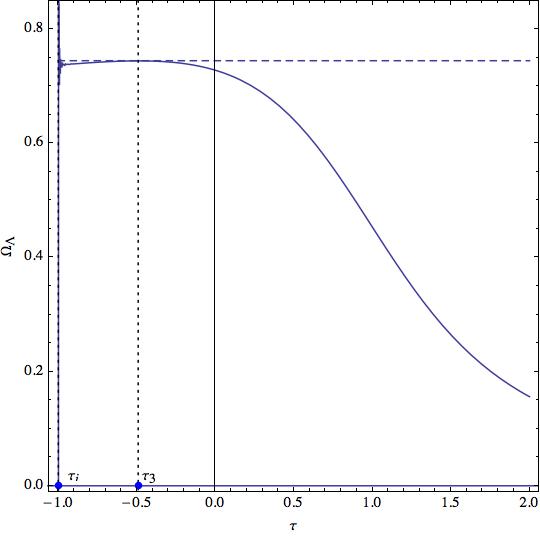}
\caption{A typical example for $\OL$ as a function of $\ta$ for $t_U\sim13.9$ Gyears and $w_0\sim - 0.962$. The dashed horizontal line is $\OL={\OL}_c$  The asymptotic continuations for $\ta\ra\in$ is $\OL\ra 1/(3K_+ \ta^2)\sim 0.718/\ta^2$.  Notice the oscillations of $\OL$ where $\ta\ra\ta_i$. This is a Mathematica feature due to the fact that $\OL$ is equal to a difference of two quantities both of which become equal but infinite. Perhaps  ${\OL}_i\ra 0$ brutally. } 
\label{Figure 2}
\end{center}
\end{figure}
%--------------------------
Perhaps the most interesting feature  of the model    is that reasonable configurations, that is for $\rh\ge0$, exist only for a limited range of values of $w_0$ which depends on $t_U$ as shown in Figure 3:
%------------Figure 3-------------------
 \begin{figure}[htbp]
\begin{center}
\includegraphics[width=8cm]{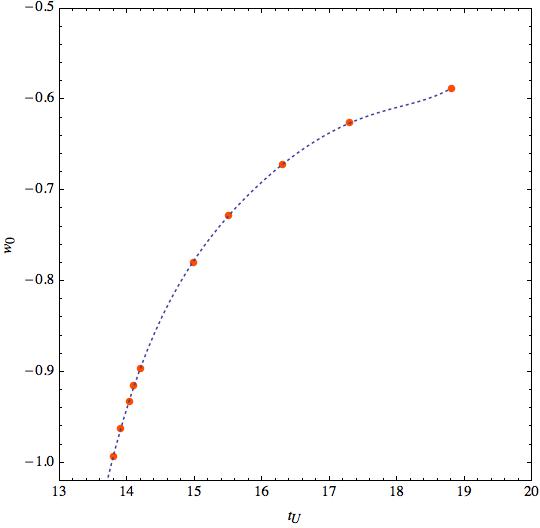}
\caption{The time derivative today ${\OL}'_0$ or rather its associated parameter $w_0$  as a function of the age of the universe $t_{U}$  in Gyears. Present day measurement is about the last point on the bottom left. The continuous dotted  line is a polynomial fit of order $5$.}  
\label{Figure 3}
\end{center}
\end{figure}
%--------------------
\ba
  ~~~~ 13.8\ls t_U\ls29.5 ~~~~,~~~ - 0.993\ls w_0\ls - 0.535.\nn
\lb{21}
\ea
Accordingly at present\footnote{The upper limit of $t_U$ grows very fast beyond $29.5$
 for very small increments of the parameters but the exact limit is hard to calculate and of little relevance.}, with $t_{U0}\sim 13.75$ Gyears the prediction is  $w_0\sim -0.99$. 
\vs
% ------------------------(ii) Problems----------------------
\nnn{\it (i) Problems with the model}
\vs
One problem is that $c_1(\ta_2)$ and $c_2(\ta_2)$ go to zero  at $\ta_1<\ta_i$ as can be seen in Figure 4. 
%------------Figure 4-------------------
 \begin{figure}[htbp]
\begin{center}
\includegraphics[width=8cm]{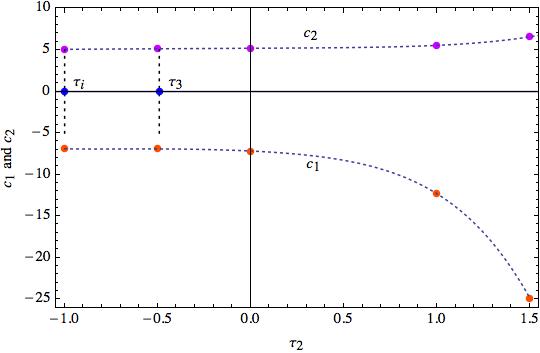}
\caption{$c_1$ and $c_2$ as functions of $\ta_2$. The dotted lines are polynomial fittings of order $5$.} 
\label{Figure 4}
\end{center}
\end{figure}
%--------------------
Unfortunately, Mathematica cannot reach beyond the singularity. This may not be a serious    flaw from a physical point of view.   
%begin changes
  Given $\ta_1$ and  $h(\ta_1)$, \eee  (\ref{15}) has  smooth numerical solutions. However, none were found with the energy density $\rh$ always positive. Another problem is that \eeee (\ref{12}) are no more Einstein's equations.
%end of changes
 It is plausible that Einstein's \eeee with their purely local conservation of the matter energy momentum tensor are not valid at cosmological scales. 
\vs
% ------------------------(iii) Conclusion----------------------
\nnn{\it (iii) Conclusion}
\vs
Equations (\ref{12}) are   in some respects appealing. Results are surprisingly close to observations with such a primitive model. No   adjustable      parameters\footnote{The assumption that   $\ka$ is constant even at $\ta_i$ may not be true.}, no scalar fields coming from nowhere, no ``quintessence".   The cosmological ``constant"     varies   mildly, is   positive and remains   small. Both $\OL$ 
and $\Om_m$ become and stay of the same order of magnitude at any later time and tend to zero in the future. The  coincidence problem is  far less acute.   The model predicts a relation between $w_0$ and   $t_{U0}$  which is remarkably close to  observations. 
For   those reasons, the present unusual variational principle deserved   some  attention.

  Finally a referee of the editorial board brought to our attention a paper by Kaloper and Padilla \cite{KP} whose motivations are light years away from ours but whose starting point  involves as here the 4-volume of \sss time  to remove the disturbing effect of the vacuum energy on the cosmological constant. The authors need a finite space   which leads    inevitably to a final crunch. We took  \sss to be flat for simplicity. A closed \sss  would certainly be more in tune with Mach's principles.

%_________-------------------------------Acknowledgments---------------------------
\vs
   {\bf  Acknowledgments}
 \vs 
 I  thank Nathalie Deruelle for her insight  into   the properties of equation (\ref{17}) which is much appreciated. I also thank Donald Lynden-Bell for his enthusiasm with the differential equations which was illuminating in many ways and very helpful. 
\vs

\end{document}